\documentstyle[12pt]{article}
\begin{document}
\addtolength{\baselineskip}{.5 \baselineskip}

\begin{flushright}
UPRF-94-399
\end{flushright}

\begin{center}
{ \LARGE \bf   Eigenvalues of the Weyl Operator \\
\vskip.2cm
as Observables of General Relativity} \\
\vskip.5cm
Roberto De Pietri\\[2 mm]
{\it Department of Physics and Astronomy, University of Pittsburgh,}\\
{\it and INFN sezione di Milano Gruppo Collegato di Parma}\\[4 mm]
and \\[4 mm]
Carlo Rovelli\footnote{On leave
from Dipartimento di Fisica - Universit\`a di Trento
and I.N.F.N., Sezione di Trento} \\
\vskip.2cm
{\it Department of Physics and Astronomy, University of Pittsburgh,}\\
{\it  Pittsburgh, Pa 15260, USA }
\end{center}

\vskip 1cm
\centerline{\bf Abstract}
\begin{quote}
We consider the eigenvalues of the three-dimensional Weyl
operator defined in terms of the (Euclidean)
Ashtekar variables, and we study
their dependence on the gravitational field. We notice
that these eigenvalues can be used as gravitational variables, and
derive explicit formulas for their Poisson brackets and their
time evolution.
\end{quote}
\noindent
P.A.C.S. number:+04.20,04.90

\newpage

A longstanding open problem in general relativity is
the problem of finding
a complete set of diffeomorphism invariant quantities;
or
quantities that have vanishing Poisson brackets with the
canonical
constraints (to our knowledge, the problem was first
posed by
Peter Bergman in reference \cite{peter}).  Quantities of
these kind
could be related to physical measurements quite
directly, and
might be effective tools for studying Einstein's
equations and for
the quantization of the theory \cite{problem}.
Considerable effort
has been expended in the past \cite{past}, and recently
\cite{RN},
on this problem.   Alain Connes has recently observed
that a
natural set of diffeomorphism invariant objects is
provided by the
spectrum of the Dirac operator of the metric manifold,
and has
suggested that the operator's eigenvalues might play a
role as
natural gravitational variables \cite{alain}.
We begin here a
simple preliminary exploration of this suggestion.

We consider the spectrum of the three-dimensional
self-dual
Weyl operator constructed in terms of the Ashtekar
formalism.
This operator, and thus its spectrum, depends on the
geometry, namely on the (Ashtekar's) gravitational
fields
$A_a(x)$ and $\tilde\sigma^a(x)$.  We can think at the
Weyl
operator's eigenvalues $\lambda_n$ as diffeomorphism
invariant
variables describing the geometry of the spacetime
manifold; or,
equivalently, as functions
$\lambda_n[A_a,\tilde\sigma^a]$ on the
phase space of general relativity, which commute with
the
diffeomorphism constraint and with the internal gauge
constraint.
In order to find the explicit form of these functions,
we should
solve the Weyl operator eigenvalue problem on a
generical
geometry, a task clearly beyond our capacities.  In
spite of this, a
surprising amount of progress can be made in analysing
the
properties of the $\lambda_n$ variables.  In particular,
we show
in this paper that their evolution equation in
coordinate time, as
well as the Poisson brackets between them can be
obtained
explicitly.

Ideally, one would like to re-express general relativity
entirely in
terms of the $\lambda_n$ variables.  This would yield an
interesting result: a fully 3-d diffeomorphism invariant
formulation of the theory.  In practice, we are far from
such a
result, because we have no control on the injectivity
and
surjectivity of the map that sends $[A_a,
\tilde\sigma^a]$ in
$\lambda_n$.  More seriously, we are not able to express
the
formulas we obtain for the Poisson brackets and for the
time
evolution of the $\lambda_n$, solely in terms of the
eigenvalues
themselves.   In spite of these clear limits, however,
we view our
preliminary results as somehow encouraging.  The time
evolution equation that we obtain is very simple,
and the theory seems to be easily adapted to a
formulation in terms of eigenvalues.  There are several
directions
in which one may proceed further. For instance, we have
not
explored the possibility of studying a four-dimensional
Dirac-like
operator.  More substantially,  operator-algebraic
techniques may
exploit the idea of describing the geometry in terms of
the
spectrum of a Dirac-like operator \cite{connes} with
much more
powerful instruments, perhaps even sidestepping the
surjectivity
problem by exploring the possibility of non-commutative
extensions of the theory \cite{connes}.  From this last
point of
view, our work can be seen as a preliminary explorations
of an
important structure underlying general relativity, which
may
deserve substantial consideration.

\vskip 1cm

We begin by considering the canonical formulations of
Euclidean
general relativity in the Astekar formalism \cite{EGR}.
This is
given as follows.  We fix a three dimensional compact
manifold M
and two real fields $A_a^i(x)$  and $\tilde{E}^a_i(x)$
on M.  We use
$a,b,..=1,2,3$ for (abstract) spatial indices and
$i,j,..=1,2,3$ for
internal su(2) indices.  We indicate coordinates on M
with $x$.
As is well known, the relation between these fields and
the
conventional metric gravitational variables is as
follows: the
Ashtekar connection $A_a^i(x)$ is the projection on a
constant
time surface of the selfdual part of the gravitational
spin
connection; its conjugate momentum $\tilde{E}^a_i(x)$ is
the
(densitized) inverse triad, which is related to the
three
dimensional metric $g_{ab}(x)$ of the constant time
surface
(which raises and lowers the spatial $a, b$ indices)  by
\begin{eqnarray}
		g\ g^{ab} = \tilde{E}^a_i\tilde{E}^b_i,
\end{eqnarray}
where $g$ is the determinant of $g_{ab}$.  We recall
that
taking $A_a^i(x)$  and $\tilde{E}^a_j(x)$ as {\it real\
} fields yields
the Euclidean theory. We have chosen a {\it compact\ }
three-space in order to simplify the spectral properties
of
the Weyl operator.

We shall use a spinorial formalism, which is more
appropriate for
dealing with the Weyl operator. The spinorial version of
the
Astekar variables is given in terms of the Pauli
matrices ${\tau_i},
i=1,2,3$, by
\begin{eqnarray}
   \tilde\sigma^{a}(x) &=& - \frac{i}{\sqrt{2}} \
\tilde{E}^a_i(x) \
{\tau_i}, \\
   A_{a}(x) &=& - \frac{i}{{2}}\  {A}_a^i(x)\ {\tau_i}.
\end{eqnarray}
Thus, $A_a(x)$ and $\tilde\sigma^a(x)$ are 2x2 complex
matrices.
In some equations we will need to write the matrix
indices
explicitly: we use upper case indices $A, B = 1,2$ for
the spinor
space on which the Pauli matrices act. Thus, the
components of the
gravitational fields are $A_a{}^A{}_B(x)$ and
$\tilde\sigma^a{}^A{}_B(x)$.  We indicate the curvature
of $A_a$
as $F_{ab}$.  We shall need also the non-densitized
version of the
soldering form
\begin{eqnarray}
   \sigma_{a} &=&  g^{-1/2}\   \tilde\sigma_{a}.
\end{eqnarray}
In terms of this notation the dynamics of Euclidean
general relativity is given by the fundamental Poisson
brackets
\begin{eqnarray}
\{A_{a}^{AB}(x), \tilde\sigma^{b}_{CD}(y) \} =
    \frac{1}{\sqrt{2}}\ \delta^b_a\  \delta^{(A}_C
\delta^{B)}_D\
\delta^3(x, y);
\label{PB}
\end{eqnarray}
and the constraints
\begin{eqnarray}
 C_b &=& \sqrt{2}\ {\rm Tr}[~\tilde\sigma^{a}F_{ab}~]
\simeq
0 , \label{SCa} \\
{\bf C} &=&  \sqrt{2} \ {\cal D}_a  \tilde\sigma^{a}
\simeq 0 ,\label{SCi} \\
S  &=&  {\rm Tr}[~\tilde\sigma^{a}
\tilde\sigma^{b}F_{ab}~]
\simeq 0. \label{SC}
\end{eqnarray}
Here $\bf C$ is a matrix in spinor space, and $S$ is a
scalar
density. These constraints generate the infinitesimal
 gauge
transformations
\begin{eqnarray}
 \delta_{N^a} f &=& \int d^3 y \left[  ~\{ f , C_a(y) \}
                                         -~\mbox{Tr}[\{
f , C(y) \}  A_{a}(y)]
                                  \right] ~ N^a(y), \\
\delta_{\bf \rho} f &=&
     \int d^3 y ~\mbox{Tr}[\{ f , {\bf C} (y) \} ~{\bf
\rho}(y)] , \\
 \delta_{\rlap{\lower1ex\hbox{$\sim$}}{N}} f &=& \int
d^3y ~\{ f ,
C(y) \} ~ \rlap{\lower1ex\hbox{$\sim$}}{N}(y) ;
\end{eqnarray}
where $N^a$ is the Shift function, which generates
spatial
diffeomorphisms; $\rlap{\lower1ex\hbox{$\sim$}}{N}$ is
the
Lapse function, which generates coordinate time
evolution; and
$\bf \rho$ is a matrix field in spinor space, which
generates the
local internal rotations.  The action of these
generators
on the fundamental variables is
\begin{eqnarray}
\delta_{\rlap{\lower1ex\hbox{$\sim$}}{N}}
\tilde\sigma^{a} &=&  \frac{1}{\sqrt{2}}
         \left[
         {\cal D}_b \left(
\rlap{\lower1ex\hbox{$\sim$}}{N}
\tilde\sigma^{b}\tilde\sigma^{a} -
\rlap{\lower1ex\hbox{$\sim$}}{N}  \tilde\sigma^{a}
\tilde\sigma^{b}
                    \right) \right] , \\
\delta_{\rlap{\lower1ex\hbox{$\sim$}}{N}}  A_{a} &=&
\frac{1}{\sqrt{2}} \left[
          \rlap{\lower1ex\hbox{$\sim$}}{T}
\tilde\sigma^{b}  F_{ab}
-  F_{ab}  \tilde\sigma^{b}
        \right],  \\
\delta_{\bf\rho} \tilde\sigma^{a} &=& - {\bf\rho}
\tilde\sigma^{a} + \tilde\sigma^{a}{\bf\rho},  \\
\delta_{\bf\rho}  A_{a} &=& {\cal D}_a {\bf\rho} ,   \\
\delta_{N^b}     \tilde\sigma^{a}&=& - N^b \partial_b
\tilde\sigma^{a} +  (\partial_a N^b)
\tilde\sigma^{a},  \\
\delta_{N^b}     A_{a} &=& - N^b \partial_b
A_{a} - (\partial_a N^b) \partial_b  A_{a} .
\label{action}
\end{eqnarray}
The problem that we consider in this work is the
definition of
gauge invariant quantities.  More precisely, we are
interested in
three-dimensional and four-dimensional observables. We
denote a
functional of the elementary fields $ F[A_a,
\tilde\sigma^{a}]$ as
{\it three-dimensional
observable\ } if
\begin{eqnarray}
 \delta_{N^b} F[A_a,  \tilde\sigma^{a}]     \simeq
 \delta_{\bf\rho} F[ A_a, \tilde\sigma^{a}] \simeq 0 ;
\end{eqnarray}
and we denote it as {\it four-dimensional observable\ }
if it is
three-dimensional observable and
\begin{eqnarray}
\delta_{\rlap{\lower1ex\hbox{$\sim$}}{N}}
F[A_a, \tilde\sigma^{a}]     \simeq 0.
\end{eqnarray}

\vskip1cm

Consider spinor fields on M, that is, two-components
complex
fields $\lambda$, with components $\lambda^A(x)$. We
follow the
standard convention \cite{PENROSE} of raising and
lowering spinor
indices as in $ \lambda_A = \lambda^{B} \epsilon_{BA}$
and
$\lambda^A = \epsilon^{AB} \lambda_B $, where
$\epsilon^{AB}$ and $\epsilon_{AB}$ are antisymmetric
and
$\epsilon_{12} = \epsilon^{12} = 1$.   The spinors in a
point
$x$ in M form a two-dimensional complex space. We
consider the
scalar product defined on this space by
\begin{eqnarray}
	(\lambda(x), \eta(x)) \equiv
\overline{\lambda}^A(x)\,
\delta_{AB}\, 	\eta^B(x) \equiv
\overline{\lambda}{}^1(x)\,
\eta^1(x) + \overline{\lambda}{}^2(x)\, \eta^2(x).
\end{eqnarray}
We use also the Ashtekar's {\it dagger} notation in
order to
indicate this scalar product:
\begin{eqnarray}
  (\eta^\dagger)^1(x) &\equiv& (\bar\eta)^2(x),
\nonumber\\
  (\eta^\dagger)^2(x) &\equiv& - (\bar\eta)^1(x) ;
\end{eqnarray}
in terms of which
\begin{eqnarray}
	(\lambda(x), \eta(x)) = (\lambda^\dagger)^A(x)
\epsilon_{AB} \eta^B(x).
\end{eqnarray}
Given the triad field $\tilde\sigma^a$ on M, we can
construct a
volume form, and therefore a scalar product on the space
of the
spinor fields $\lambda$. This is given by
\begin{eqnarray}
\langle\lambda,\eta\rangle \equiv \int d^3x
{}~\sqrt{g(x)}\
(\lambda(x), \eta(x)).
\end{eqnarray}
We write this also as
\begin{eqnarray}
\langle\lambda,\eta\rangle \equiv \int_M \  \lambda
\cdot \eta.
\end{eqnarray}
where the volume form defined by the triad and the
scalar
product are understood.  Equipped with the product
$\langle\ ,\
\rangle $, the spinor fields form an Hilbert space $\cal
H$.
The spinorial Ashtekar connection naturally defines an
SU(2)
covariant derivative on the spinor fields
\begin{eqnarray}
 {\cal D}_a \eta &=& \partial_a \eta  +  A_{a}\, \eta,
\end{eqnarray}
where $A_a$ acts on $\eta$ by matrix multiplication. We
also
recall that the Ashtekar connection can be decomposed as
\begin{eqnarray}
  A_{a} &=& {\mbox{\boldmath $\Gamma$}}_a [ \sigma{}] -
\frac{1}{\sqrt{2}} {\Pi_a},
\label{DA}
\end{eqnarray}
where ${\mbox{\boldmath $\Gamma$}}_a$ is the (unique)
symmetric spinor connection compatible with the SU(2)
soldering
$\sigma_a$; and on the constraints  surface we have
${\Pi_a} =
K_{ab} \sigma^b$ where $K_{ab}$ is the extrinsic
curvature of the
three manifold.

Let us now introduce the main object we will deal with.
We
consider the operator $\hat{H}$ defined as
\begin{eqnarray}
\hat{H}\ {\eta}  &=& \sqrt{2}\  \sigma^{a}\ {\cal D}_a
\eta.
\end{eqnarray}
The operator $\hat{H}$ is the 3-dimensional self-dual
Weyl
operator; it has a precise physical interpretation, as
the operator
that generates the dynamics of a left-handed massless
fermion, as
a neutrino, on a gravitational background.  Indeed,  a
massless
spinor test-particle satisfies
\cite{WEIL}
\begin{eqnarray}
  \frac{d}{dt}\, \eta(x,t) &=& \hat{H}\, \eta(x,t).
\label{ferm}
\end{eqnarray}
We will not make any explicit use here of this dynamical
interpretation of the Weyl operator $\hat H$.

The Weyl operator is symmetric in $\cal H$, that is
\begin{eqnarray}
\langle\hat{H}\lambda,\eta\rangle =
\langle\lambda,\hat{H}\eta\rangle .
\end{eqnarray}
This can be proven by integrating by part, using the
reality of the
fields $A_a^i$ and $E^a_i$ and the hermiticity
properties of the
Pauli matrices,  and  making use of the constraint $\bf
C$
(equation \ref{SCi}).  Neglecting domain's difficulties,
we assume
that $\hat{H}$ is self-adjoint.   Then we can consider
its
eigenvalue problem
\begin{eqnarray}
      \hat{H}\  \eta_n  &=&  \lambda_n \ \eta_n.
\end{eqnarray}
The eigenspinors $\eta_n$ form an orthonormal basis of
(possibly
generalized) vectors in $\cal H$.  We assume that they
form a
countable basis (this is not unreasonable, since M is
compact),
and we take the indices $n$ as integers (see below).
The eigenvalues $\lambda_n$
are the objects on which we focus. Since $\hat H$
depends on the gravitational variables $A_a$ and
$\tilde\sigma^a$, so do its eigenvalues. Thus, the
eigenvalue
equation implicitly defines a countable family of
functionals of
the gravitational fields
\begin{eqnarray}
		\lambda_n = \lambda_n[ \tilde\sigma^{a},
A_{a}].
\label{functionals}
\end{eqnarray}
The eigenvalues $\lambda_n$ can thus be seen as a family
of
variables describing the gravitational field.

The functionals (\ref{functionals}) are determined by the
eigenvalue equation only implicitly. To find their
explicit form
we should solve the spectral problem for the operator
$\hat H$ on
arbitrary geometries. We can nevertheless obtain a
large amount of important information about these
functionals.  In
particular, we can compute their first variation, namely
their
derivative with respect to the gravitational variables
$A_a$ and
$\tilde\sigma^a$ .   This allows us to check their gauge
invariance
explicitly and, more importantly, to write explicit
formulas for
their Poisson brackets and their time evolution. These
are
substantial steps towards the task of expressing
general relativity
entirely in terms of the $\lambda_n$ variables alone.
The
computation of the first variation of the $\lambda_n$'s
is
an application of the technology of quantum mechanics'
time-independent perturbation theory.  Consider an
operator
$\hat O$ with a complete set of eigenstates  $v_n$ and
eigenvalues $\lambda_n$
\begin{eqnarray}
		\hat O\ v_n = \lambda_n \ v_n.
\end{eqnarray}
Let $\hat O$, and thus its eigenvectors and eigenvalues,
depend
on a parameter $\tau$.  We want to compute $d\lambda_n /
d\tau$. To this aim, consider a small variation
$\tau\rightarrow
\tau+\delta\tau$, which induces the variations $\delta
O, \delta
v_n $ and $\delta \lambda_n$.  The variation of the
eigenvalue
equation gives
\begin{eqnarray}
	\delta\hat O\, v_n + \hat O \, \delta v_n =
\delta \lambda_n
\, v_n + \lambda_n \, \delta v_n.
\end{eqnarray}
By taking the scalar product of this equation with the
eigenstate
$v_n$  (on the left), and using the orthogonality
property and the
self-adjointness of $\hat O$, we obtain the formula,
well known
from quantum mechanics,
\begin{eqnarray}
	\delta \lambda_n = (v_n, \delta \hat O\, v_n).
\label{variatio}
\end{eqnarray}

Let us apply this to the Weyl operator. Consider a small
variation
$A_{a}(x) \rightarrow  A_{a}(x) + \delta  A_{a}(x)$ and
$
\tilde\sigma^{a} (x) \rightarrow  \tilde\sigma^{a}(x) +
\delta
\tilde\sigma^{a}(x)$. From the definition of the Weyl
operator we
have immediately
\begin{eqnarray}
	{\delta \hat H \over \delta A_a(x)} = \sqrt{2}  \
\tilde\sigma^{a}(x)
\end{eqnarray}
and
\begin{eqnarray}
	{\delta \hat H \over \delta  \sigma^{a}(x)} =
\sqrt{2}\ {\cal
D}_a.
\end{eqnarray}
We need the derivative with respect to the {\it
densitized} triad.
This is obtained using
\begin{eqnarray}
 \delta \sigma^{a} = \delta (g^{-1/2} \tilde\sigma^{a})
&=&
  g^{-1/2}\delta  \tilde\sigma^{a} + \frac{1}{2}
\tilde\sigma^{a}
\mbox{Tr} [ g^{-1} \tilde\sigma_{c} \delta
\tilde\sigma^{c}]
\end{eqnarray}
By inserting these equations in the relation
(\ref{variatio}) for the
Weyl operator, we obtain with some simple algebra the
key
technical result
\begin{eqnarray}
 \delta \lambda_n &=& - \sqrt{2} \int_M
    \left[ \eta_n{}\cdot \delta\tilde\sigma^{a}{\cal
D}_a \eta_n  +
\frac{1}{2} \eta_n  \cdot \mbox{Tr}[\sigma_{a} \delta
\tilde\sigma^{a}] \ \sigma^{b}{\cal D}_b \eta_n \right]
\nonumber\\
&&  - \sqrt{2} \int_M \eta_n \cdot
\tilde\sigma^{a}~\delta A_{a}\eta_n.
\label{delta}
\end{eqnarray}
This equation allows us to compute explicitly the
variation of the
eigenvalues under arbitrary variations of the Ashtekar
variables.
Equivalently, we can write the functional derivatives
\begin{eqnarray}
 {\delta \lambda_n\over\delta\tilde\sigma^a{}^A_B}
&=& - \sqrt{2}
    \left[ \eta^\dagger_n{}^B  ({\cal D}_a \eta_n)_A  +
\frac{1}{2} \eta^\dagger_n{}^C \ \sigma^{b}{}_{CD} \,
           ({\cal D}_b \eta_n)^D \sigma_{a}{}^B_{~A}
\right]
\label{d1}\\
 {\delta \lambda_n\over\delta A_a{}^A_B}
&= &  - \sqrt{2} \eta^\dagger_n{}^C
\tilde\sigma^{a}{}_{CA}~\eta_n^B.
\label{d2}
\end{eqnarray}

As a first use of equation (\ref{delta}), let us check
the invariance
of the eigenvalues under diffeomorphisms and SU(2)
transformations explicitly.  By
replacing $\delta \tilde\sigma^{a}$ and $\delta A_{a}$
with the
gauge variations given in equation (\ref{action}), we
have with
simple algebra
\begin{eqnarray}
 \delta_{\bf\rho} \lambda_n &=& 0, \\
 \delta_{N^a} \lambda_n &=& 0;
\end{eqnarray}
namely the eigenvalues are gauge invariant under
three-dimensional diffeomorphisms and internal SU(2)
gauge
transformations.

By substituting the variations $A_{a}$ and $
\tilde\sigma{a}$
generated by the Hamiltonian constraint in equation
(\ref{delta}),
we obtain the formula for the evolution of the
eigenvalues in
coordinate time, with an arbitrary Lapse function
$\rlap{\lower1ex\hbox{$\sim$}}{N}$
\begin{eqnarray}
 \delta_{\rlap{\lower1ex\hbox{$\sim$}{}}{N}} \lambda_n
&=&
   \sqrt{2} \int d^3\! x ~\sqrt{g}\
     \eta^{\dagger A}_n \epsilon_{AB}
     \left[ (\partial_a N) \epsilon^{abc}
\sigma_{c}{}^B_{~C}
           + K^b_{~a}  \sigma^{a}{}^B_{~C}
     \right]  ~{\cal D}_b \eta^C_n.
\end{eqnarray}
If we chose a spatially constant Lapse function, the
time evolution
of the eigenvalues (which we can indicate now as
$\dot\lambda\equiv d/dt\, \lambda$) is given by
\begin{eqnarray}
 \dot\lambda_n =\sqrt{2} \int_M \eta_n \cdot K^a\,{\cal
D}_a
\eta_n.
\label{evolution}
\end{eqnarray}
This is our main result.

Similarly, a straightforward calculation yields the
Poisson brackets
between two $\lambda$'s. From the definition of the
bracket
\begin{eqnarray}
\{ \lambda_n , \lambda_m \} = \int d^3x\
{\delta\lambda_n\over
\delta A_a(x)} {\delta\lambda_m\over \delta
\tilde\sigma^a(x)} -
{\delta\lambda_n\over \delta \tilde\sigma^a(x)}
{\delta\lambda_m\over \delta A_a(x)}
\end{eqnarray}
we obtain by inserting the derivatives (\ref{d1}) and
(\ref{d2}),
and with some simple algebra
\begin{eqnarray}
\{ \lambda_n , \lambda_m \} &=&
  \frac{1}{2} (\lambda_n - \lambda_m )
  \int d^3\!x \sqrt{q}                  \\
&&~~~
\bigg[
     ({\eta_m}^\dagger_A {\eta_n}^A)~({\eta_n}^\dagger_B
{\eta_m}^B)
     -\frac{3}{2}  ({\eta_n}^\dagger_A {\eta_n}^A)~
({\eta_m}^\dagger_B {\eta_m}^B) \bigg]
    \nonumber \\
&& + \frac{1}{2} \int d^3\!x \sqrt{q}
  ( {\eta_n}^{\dagger A}   \sigma^{a}_{AB}
{\eta_m}^{\dagger B})
\left[{\eta_m}_C {\cal D}_a {\eta_n}^C
      -{\eta_n}_C {\cal D}_a {\eta_m}^C
        \right] \nonumber
\label{Poisson}
\end{eqnarray}
A final property of the eigenvalues that we note is that
if
$\lambda$ is an eigenvalue of the Weyl operator, so is
$-
\lambda$. In fact, it is easy to verify that
\begin{eqnarray}
 \hat{H} \eta = \lambda \eta
\end{eqnarray}
implies
\begin{eqnarray}
\hat{H} \eta^{\dagger}= -\lambda \eta^{\dagger}
\end{eqnarray}
We can therefore choose $\lambda_{-n}=-\lambda_n$.

\vskip1cm

As indicated above, we would like to
consider the
map $(A_a,\tilde\sigma^a) \rightarrow\lambda_n[A_a,
\tilde\sigma^a]$ as a change of variables on the phase
space of
general relativity.  Given the dynamical interpretation
of the Weyl
operator (see equation \ref{ferm}), the variables
$\lambda_n$
are related to the energy spectrum of the massless
left-handed fermions.  They capture geometrical features
of the
spacetime manifold via dynamical properties of these
test fields.
Many problems, however, remain open.
We have little control on
the properties of this change of variables.
First, we do not know
the constraints satisfied by the $\lambda_n$ variables;
namely
the properties that a generic set of numbers $\lambda_n$
should
have in order to be interpretable as the spectrum of a
Weyl
operator (surjectivity).  Second, the other way around,
it would be
important to show that generically the map is
non-degenerate;
namely that the vanishing of the right hand side of
equation
(\ref{delta}) implies (generically) that the variations
of $A_a$ and
$\tilde\sigma^a$ are the gauge variations
$\delta_{\bf\rho}$ and
$ \delta_{N^a}$ given in (\ref{action}) (injectivity).
We suspect this
is true, but we have not been able to prove it.  If the
injectivity of
the map could be proven, we would be assured that the
map can
be inverted, and therefore that the right hand side of
equations
(\ref{evolution}) and (\ref{Poisson}), which represent
our
main results,  can be expressed solely in
terms of the $\lambda_n$ variables.  If this could be
done
explicitly we would have a three-dimensional
diffeomorphism-invariant formulation of general
relativity.

\vskip1cm

The idea of this work has emerged from a converastion with
Chris Isham and Alain Connes. We are indebited with
both of them;  and with the Isaac Newton Institute in
Cambridge, where, and thanks to which,
this conversation has taken place.
This work was partially supported by the NSF grant
PHY-9311465, and by the INFN ``Iniziativa Specifica FI2''.

\end{document}